\documentclass[showpacs,preprintnumbers,amsmath,amssymb,showkeys,nofootinbib,12pt]{revtex4}
\usepackage[dvips]{graphicx}% Include figure files
\usepackage{dcolumn}% Align table columns on decimal point
\usepackage{bm}% bold math
\usepackage{latexsym}
\usepackage{amsfonts}
\usepackage{amssymb}
\usepackage{amsmath}
\usepackage[colorlinks]{hyperref}
\begin{document}

\title{Logarithmic entropy--corrected holographic dark energy with non--minimal kinetic coupling}
\author{Ali R. Amani$^{1}$}
\thanks{Corresponding Author:
a.r.amani@iauamol.ac.ir} \affiliation{$^{3}$Islamic Azad University,
Ayatollah Amoli Branch, Department of Physics,\\ P.O. Box 678, Amol,
Mazandaran, Iran}

\author{J. Sadeghi$^{1}$}\thanks{Email:
pouriya@ipm.ir} \affiliation{$^{3}$Islamic Azad University,
Ayatollah Amoli Branch, Department of Physics,\\ P.O. Box 678, Amol,
Mazandaran, Iran}

\author{H. Farajollahi$^{4}$}\thanks{Email: hosseinf@guilan.ac.ir}
\author{M. Pourali$^{4}$}
\thanks{Email: m.pourali@msc.guilan.ac.ir}

 \affiliation{$^{4}$Department of Physics, University of Guilan,
Rasht, Iran}

\date{\today}



\begin{abstract}
In this paper,  we  have considered a cosmological model with the non--minimal kinetic coupling
terms and investigated its cosmological implications with respect to the logarithmic entropy--
corrected holographic dark energy (LECHDE). The
correspondence between LECHDE in flat FRW cosmology and the phantom dark
energy model with the aim to interpret the current universe acceleration is also examined.
\end{abstract}
\maketitle

\section{Introduction}

The universe acceleration, shown
by several astronomical observations, indicates
the existence of a mysterious exotic matter called dark energy (DE) \cite{Riess}.
In the classical gravity,  whereas the cosmological constant, $\Lambda$, is the most
prominent candidate of DE with the equation of state (EoS) parameter equals -1 \cite{Weinberg},
there are strong evidence for a dynamical DE equation of state. However, the problem of DE \cite{Setare},
its energy density and EoS parameter is still an unsolved problem in classical gravity
and may be in the context of quantum gravity we achieve a more inclusive insight to its properties \cite{Cohen}.

While in the microscopic level, the Einstein's theory of gravity still
remains unclear, the authors in \cite{J.M,J.D,S.W,T.J} obtain the first
law of thermodynamics for black holes. Later, Padmanabhan
 proposes a thermodynamic interpretation of gravity
\cite{T.P1,T.P2}. Recently, explanation of gravity as entropic
force is pointed out by Verlinde \cite{E.P} in several models such as specific microscopic model
of space-time \cite{J.M1}, construction of holography from
black hole entropy \cite{F.C} and quantum information theories \cite{J-W}. In addition, a
modified entropic force in the Debye model is presented in Ref.
\cite{C-J}. For other relevant works in entropic force, see Refs.
\cite{J.M2,Yu-X,Jao-L,I.V.V,Y.Ti,yun S,Ah sh}. In particular, the
holographic principle is discussed in details by several authors
\cite{F.W.Sh,R.G.C,y.z.g,S.W.Wei,Y.L.G.P,SJM}. In \cite{T1W}, it has been shown
that the holographic dark energy can be derived from the entropic
force. The correspondence between entropy--corrected
holographic and Gauss--Bonnet dark energy models is discussed is \cite{SeJa}.
Here, we intend to investigate the
correspondence between logarithmic entropy--corrected holographic and cosmological
models with kinetic terms coupled non--minimally
to the scalar field and to the curvature \cite{L. N. Granda1,L. N. Granda2,L. N. Granda3}
as a source of
dark energy \cite{L. N. Granda1,L. N. Granda2,L. N. Granda3}. The
initial motivation to study such theories is related to low energy
limit of several higher dimensional theories.

The paper is organized as follows:
\par In section \ref{2}, we review the scalar-tensor theories with
non--minimally coupled kinetic terms to the curvature and the scalar
field. We will obtain the field equations and energy--momentum
tensors. In section \ref{3} we introduce the basic setup of the
ECHDE and obtain the corresponding EoS parameter for further
investigation. In section \ref{4}, the correspondence between ECHDE
and non-minimal kinetic coupling term and acceleration of the
universe are presented.  A short summary is given in section
\ref{5}.
%$$$$$$$$$$$$$$$$$$$$$$$$$$$$$$$$$$$$$$$$$$$$$$$$$$$$$$$$$$$$$$$$$$$$$$$$$$$
\section{The model}\label{2}
The standard FRW cosmological is given by the metric,
\begin{equation}\label{ds2}
 ds^2 = -dt^2 + a(t)^2 \left(
{dr^2 \over 1- kr^2} + r^2 ( d\theta^2 + \sin^2\theta d\phi^2)
\right),
\end{equation}
where $a(t)$ is scale factor and $k=0,+1,-1$ imply the flat, close
and open universe respectively. The energy-momentum tensor of a
perfect fluid is given by $T^\mu_\nu=$ diag($-\rho,p,p,p)$. The
Friedmann equations then are,
\begin{equation}\label{fried1}
\begin{aligned}
   & H^2=\frac{\kappa^2}{3} \rho , \\
    &\dot{H}=-\frac{\kappa^2}{2} (\rho+p).
\end{aligned}
\end{equation}
We start with a cosmological model in which there is an interaction
between a scalar field and the curvature \cite{L. N. Granda1,L. N.
Granda2,L. N. Granda3}
\begin{equation}\label{action}
\begin{aligned}
S=&\int d^{4}x\sqrt{-g}\Big[\frac{1}{16\pi G}
R-\frac{1}{2}\partial_{\mu}\phi\partial^{\mu}\phi-
\frac{1}{2} \xi R \left(F(\phi)\partial_{\mu}\phi\partial^{\mu}\phi\right) -\\
&\frac{1}{2} \eta
R_{\mu\nu}\left(F(\phi)\partial^{\mu}\phi\partial^{\nu}\phi\right) -
V(\phi)\Big],
\end{aligned}
\end{equation}
where the coupling constants of dimensionless $\xi$ and $\eta$ are kinetic coupling and depend on the type of function $F(\phi)$.\\
 Thus, by taking variation of action (\ref{action}) with respect to the metric, we have,
\begin{equation}\label{rmu}
R_{\mu\nu}-\frac{1}{2}g_{\mu\nu}R=\kappa^2
\left[T_{\mu\nu}^{\phi}+T_{\mu\nu}^{\xi}+T_{\mu\nu}^{\eta}\right],
\end{equation}
where $\kappa^2=8\pi G$.\\
The $T_{\mu\nu}^{\phi}$ is the energy-momentum tensor for the scalar field $\phi$, and
$T_{\mu\nu}^{\xi}$ and $T_{\mu\nu}^{\eta}$ are the energy-momentum tensor for the minimally coupling $\xi$ and $\eta$ respectively. So, the corresponding energy-momentum
tensors are,
\begin{equation}\label{tphi}
T_{\mu\nu}^{\phi}=\nabla_{\mu}\phi\nabla_{\nu}\phi-\frac{1}{2}g_{\mu\nu}\nabla_{\lambda}\phi\nabla^{\lambda}\phi
-g_{\mu\nu}V(\phi),
\end{equation}
\begin{equation}\label{txi}
\begin{aligned}
T_{\mu\nu}^{\xi}=&\xi\Big[\left(R_{\mu\nu}-\frac{1}{2}g_{\mu\nu}R\right)\left(F(\phi)\nabla_{\lambda}\phi\nabla^{\lambda}\phi\right)+
g_{\mu\nu}\nabla_{\lambda}\nabla^{\lambda}\left(F(\phi)\nabla_{\gamma}\phi\nabla^{\gamma}\phi\right)\\
&-\frac{1}{2}(\nabla_{\mu}\nabla_{\nu}+\nabla_{\nu}\nabla_{\mu})\left(F(\phi)\nabla_{\lambda}
\phi\nabla^{\lambda}\phi\right)+R\left(F(\phi)\nabla_{\mu}\phi\nabla_{\nu}\phi\right)\Big],
\end{aligned}
\end{equation}
and
\begin{equation}\label{teta}
\begin{aligned}
T_{\mu\nu}^{\eta}=&\eta\Big[F(\phi)\left(R_{\mu\lambda}\nabla^{\lambda}\phi\nabla_{\nu}\phi+
R_{\nu\lambda}\nabla^{\lambda}\phi\nabla_{\mu}\phi\right)-\frac{1}{2}
g_{\mu\nu}R_{\lambda\gamma}\left(F(\phi)\nabla^{\lambda}\phi\nabla^{\gamma}\phi\right)\\
&-\frac{1}{2}\left(\nabla_{\lambda}\nabla_{\mu}\left(F(\phi)\nabla^{\lambda}\phi\nabla_{\nu}\phi\right)+
\nabla_{\lambda}\nabla_{\nu}\left(F(\phi)\nabla^{\lambda}\phi\nabla_{\mu}\phi\right)\right)\\
&+\frac{1}{2}\nabla_{\lambda}\nabla^{\lambda}\left(F(\phi)\nabla_{\mu}\phi\nabla_{\nu}\phi\right)+
\frac{1}{2}g_{\mu\nu}\nabla_{\lambda}\nabla_{\gamma}\left(F(\phi)\nabla^{\lambda}\phi\nabla^{\gamma}\phi\right)\Big]
\end{aligned}
\end{equation}
In order to obtain the equation of motion for the scalar field, we
take variation of action with respect to $\phi$, so we have,
\begin{equation}\label{teta}
\begin{aligned}
&-\frac{1}{\sqrt{-g}}\partial_{\mu}\left[\sqrt{-g}\left(\xi R
F(\phi)\partial^{\mu}\phi+\eta R^{\mu\nu}F(\phi)\partial_{\nu}\phi+
\partial^{\mu}\phi\right)\right]+\frac{dV}{d\phi}+\\
&\frac{dF}{d\phi}\left(\xi
R\partial_{\mu}\phi\partial^{\mu}\phi+\eta
R_{\mu\nu}\partial^{\mu}\phi\partial^{\nu}\phi\right)=0.
\end{aligned}
\end{equation}
For simplicity we assume that $\eta + 2\xi$ = 0. Therefore, from
Eqs. (\ref{tphi})-(\ref{teta}), the energy density, pressure and the
scalar field equation are given by,
\begin{equation}\label{rho}
    \rho=\frac{1}{2}\dot{\phi}^2+V(\phi)+9\xi H^2F(\phi)\dot{\phi}^2,
\end{equation}
\begin{equation}\label{p}
    p=\frac{1}{2}\dot{\phi}^2-V(\phi)-\xi\left(3H^2+2\dot{H}\right)F(\phi)\dot{\phi}^2-2\xi H
    \left(2F(\phi)\dot{\phi}\ddot{\phi}+\frac{dF}{d\phi}\dot{\phi}^3\right),
\end{equation}
\begin{equation}\label{eqmotion}
\ddot{\phi}+3H\dot{\phi}+\frac{dV}{d\phi}+3\xi
H^2\left(2F(\phi)\ddot{\phi}+\frac{dF}{d\phi}\dot{\phi}^2\right)
+18\xi H^3F(\phi)\dot{\phi}+12\xi H\dot{H}F(\phi)\dot{\phi}=0.
\end{equation}

%################################################################################
\section{Logarithmic entropy--corrected holographic dark energy}\label{3}

The black hole entropy plays a central role in the derivation of
holographic dark energy (HDE). Indeed, the definition and derivation
of holographic energy density depends on the entropy-area
relationship $S\sim A \sim L^2$ of black holes in Einstein's
gravity, where $A \sim L^2$ represents the area of the horizon.
However, this definition can be modified from the inclusion of
quantum effects, motivated from the loop quantum gravity (LQG). The
quantum corrections provided to the entropy-area relationship leads
to the curvature correction in the Einstein-Hilbert action and vice
versa \cite{13}. The corrected entropy takes the form \cite{14}
\begin{equation}\label{entropy}
S=\frac{A}{4}+\tilde\gamma \ln\Big(\frac{A}{4}\Big)+\tilde\beta,
\end{equation}
where $\tilde\gamma$ and $\tilde\beta$ are dimensionless constants
of order unity. The exact values of these constants are not yet
determined and are still debatable in LQC. The
corrections are due to thermal
equilibrium  and quantum fluctuations \cite{15}. The second term in (\ref{entropy})
appears in a model of entropic cosmology which unifies the inflation
and late time acceleration \cite{cai}. The $\tilde{\gamma}$
 might be extremely large due to current cosmological
constraint, which inevitably brought a fine tuning problem to
entropy corrected models and it is desirable to determine
it by observational constrain. Taking the corrected entropy-area relation
(\ref{entropy}) into account, the energy density of the HDE will be
modified as well. On this basis, Wei \cite{16} proposed the energy
density of the so-called ECHDE in the form
\begin{equation}
 \rho_{\Lambda}=3c^2R_h^{-2}+\gamma R_h^{-4}\ln(R_h^{2})+\beta
 R_h^{-4},
\label{holo}
\end{equation}
in units where $M_p^2=8\pi G=1$, and $c$ is a constant
determined by observational fit. The future event horizon $R_h$
is defined as,
\begin{equation}
R_h= a\int_t^\infty \frac{dt}{a}=a\int_a^\infty\frac{da}{Ha^2},
\label{Rh}
\end{equation}
which leads to results compatible with observations. Furthermore, we
can define the dimensionless dark energy as:
\begin{equation}
\Omega_{\Lambda}\equiv\frac{\rho_{\Lambda}}{3H^2}=\frac{3c^2+\gamma
R_h^{-2}\ln(R_h^2)+\beta R_h^{-2}}{3H^2R_h^2}\label{omega}.
\end{equation}
In the case of a dark-energy dominated universe, dark energy evolves
according to the conservation law,
\begin{equation}\label{coneq}
\dot{\rho}_{\Lambda}+3H(\rho_{\Lambda}+p_{\Lambda})=0,
\end{equation}
or equivalently
\begin{equation}
\dot{\Omega}_\Lambda=-\frac{2\dot H}{3H^3R_h^2}\left(3c^2+\gamma
R_h^{-2}\ln(R_h^{2})+\beta R_h^{-2}\right)+\frac{HR_h-1}{3H^2R_h^3}\Big[
-6c^2+2\gamma R_h^{-2}-4\gamma R_h^{-2}\ln R_h^{2}-4\beta R_h^{-3}
\Big],
\end{equation}
therefore the EoS parameter reduce to,
\begin{eqnarray}\label{index}
&w_{\Lambda }=\frac{p_{\Lambda}}{\rho_{\Lambda}}=-1-\frac{ 2\gamma
R_h^{-2} -4\gamma R_h^{-2}\ln (R_h^{2})-4\beta
R_h^{-2}-6c^2}{3(3c^2+\gamma R_h^{-2}\ln (R_h^{2})+\beta
R_h^{-2})}\left[1-\sqrt{\frac{3\Omega_\Lambda}{3c^2+\gamma
R_h^{-2}\ln(R_h^2)+\beta R_h^{-2}}}\right].
\end{eqnarray}\\
%###########################################################################
\section{Correspondence between ECHDE and Non--minimal Kinetic coupling}\label{4}
Here we obtain the conditions for correspondence between our
cosmological model with the non--minimal kinetic term and the ECHDE
scenario in the flat FRW space. This can be achieved by obtaining an
appropriate potential in the model. In the following we make two
assumptions \cite{L. N. Granda2},\\
1) the function  $F(\phi)$ is in exponential form as,
\begin{equation}\label{fphi}
F(\phi)=\frac{1}{\phi_{0}^2} e^{2 \lambda \phi},
\end{equation}
and\\
 2) the scalar factor is in power laws as  $a=a_0t^{n}$,
 \cite{NOS}. For negative $n$, the scale factor does not correspond to expanding universe
but to shrinking one. If one changes the direction of time as
$t\rightarrow -t$, the expanding universe whose scale factor is
given by $a=a_0(-t)^{n}$ emerges. We note that when $t$ arrives
$t_s$ occurs a big rip singularity, so this is an important scenario
in relation with other cosmological singularities \cite{jims,cai2}.
Since $n$ is not an integer in general, the sign of $t$ is still a
problem. To avoid the inconsistency, we may further shift the origin
of the time as $-t\rightarrow t_s-t$. Then the time $t$ can be
positive as long as $t < t_s$, and we can consistently take
$a=a_0(t_s-t)^{n}$. So that, we can finally write scalar field as
Ref. \cite{L. N. Granda2} in the following form,
\begin{equation}
\label{n}
 H=\frac{n}{t},
\hspace{1cm}\phi=\frac{1}{\lambda} \ln(
\frac{\lambda\phi_{o}t}{\kappa \sqrt{\xi(1+3n)}}),
 \end{equation}
 when
$n> 0$ or
\begin{equation}\label{h01}
 H=\frac{-n}{t_s-t}, \hspace{1cm}\phi=\frac{1}{\lambda} \ln
\frac{\lambda\phi_{0}(t_s-t)}{\kappa\sqrt{\xi(1+3n)}},
\end{equation}
when $n< 0$.
Here we  first consider $n>0$ . If we establish a
correspondence between the holographic dark energy and non--minimal
coupling approach, then by using dark energy density equation
(\ref{rho}) and relation (\ref{omega}), together with expressions
(\ref{n}), we easily  arrive to the
scalar potential as,
\begin{equation}
\label{Vphi}
V=\frac{3n^{2}\lambda^{2}\phi_{0}^{2}}{\kappa^{4}\xi(1+3n)}\left(\kappa^{2}\Omega_{\Lambda}
-\frac{3}{1+3n}\right)e^{-2\lambda\phi}
 \end{equation}
The equations (\ref{n}) help us to obtain  $t$ in terms of the
scalar field $\phi$. Therefor, by substituting (\ref{fphi}), (\ref{n}),
and (\ref{Vphi}) into (\ref{eqmotion}), finally we have following
equation,
\begin{equation}
\label{7}
 3n-6n^{2}\lambda^{2}\Omega_{\Lambda}+\frac{18n^{2}\lambda^{2}}{\kappa^{2}(1+3n)}+3n^{2}
 \lambda\Omega_{\Lambda}'
+\frac{6n^{2}\lambda^{2}}{\kappa^{2}(1+3n)}\Big(3n\phi_{0}^2+1-3\phi_{0}^{2}\Big)
e^\frac{2\lambda\phi}{\phi_{0}^2}e^{-2\lambda\phi}-1=0
\end{equation}
where
\begin{equation}
\Omega_{\Lambda}'=\frac{d\Omega_{\Lambda}}{d\phi}=\frac{d\Omega_{\Lambda}}{dt}\lambda t=\frac{d\Omega_{\Lambda}}
{dt}\frac{\kappa\sqrt{\xi(1+3n)}}{\phi_{0}}\,e^{\lambda\phi}. \label{8}
\end{equation}
In order to have finite  $R_h$, we consider the ansatz $a=a_0t^{n}$
for $n>1$ and substitute  into equation  (\ref{Rh}),
 we then find:
\begin{equation}
  R_h=\frac{t}{n-1},
\end{equation}
\begin{eqnarray}\label{omlam}
\Omega_{\Lambda}=\frac{(n-1)^2}{3n^2}\Big[3c^2+\gamma
\sigma^{-2} e^{-2 \lambda \phi}
\ln\{\sigma^{2} e^{2 \lambda \phi}\}+\beta
\sigma^{-2} e^{-2 \lambda \phi}\Big],
\end{eqnarray}
and
\begin{eqnarray}
w_{\Lambda }=-1-\frac{ 2\gamma
\sigma^{-2} e^{-2 \lambda \phi} -4\gamma
\sigma^{-2} e^{-2 \lambda \phi}\ln
(\sigma^{2} e^{2 \lambda \phi})-4\beta
\sigma^{-2} e^{-2 \lambda \phi}-6c^2}{3(3c^2+\gamma
\sigma^{-2} e^{-2 \lambda \phi}\ln
(\sigma^{2} e^{2 \lambda \phi})+\beta
\sigma^{-2} e^{-2 \lambda \phi})}\nonumber\\\times\left[1-\sqrt{\frac{3\Omega_\Lambda}{3c^2+\gamma
\sigma^{-2} e^{-2 \lambda \phi}\ln(\sigma^{2} e^{2 \lambda \phi})+\beta
\sigma^{-2} e^{-2 \lambda \phi}}}\right].\label{wh0}
\end{eqnarray}
where $\sigma=\frac{\kappa\sqrt{\xi(1+3n)}}{\lambda\phi_{0}(n-1)}$.\\
For $n<0$, we repeat the process, but impose relations (\ref{h01}).
So, we find that,
\begin{equation}\label{Vphi1}
V=\frac{3n^{2}\lambda^{2}\phi_{0}^{2}}{\kappa^{4}\xi(1+3n)}\left(\kappa^{2}\Omega_{\Lambda}
-\frac{3}{1+3n}\right)e^{-2\lambda\phi}
 \end{equation}
and
\begin{equation}\label{conb}
3n-6n^{2}\lambda^{2}\Omega_{\Lambda}
+\frac{18n^{2}\lambda^{2}}{\kappa^{2}(1+3n)}+3n^{2}\lambda\Omega_{\Lambda}'
+\frac{6n^{2}\lambda^{2}}{\kappa^{2}(1+3n)}\Big(3n\phi_{0}^2
+1-3\phi_{0}^{2}\Big)e^\frac{2\lambda\phi}{\phi_{0}^2}e^{-2\lambda\phi}-1=0,
\end{equation}
where
\begin{equation}\label{81}
\frac{d\Omega_{\Lambda}}{d\phi}=-\frac{d\Omega_{\Lambda}}{dt}\lambda(t_s-t)=-\frac{d\Omega_{\Lambda}}
{dt}\frac{\kappa\sqrt{\xi(1+3n)}}{\phi_{0}}\,e^{\lambda\phi}.
\end{equation}
Now, under the ansatz $a=a_0(t_s-t)^{n}$ we can see from (\ref{Rh})
that $R_h$ is always finite if $n<0$, which is just the case under
investigation. Then we have:
\begin{equation}
R_h=\frac{t_s-t}{1-n},
\end{equation}
\begin{eqnarray}
\Omega_{\Lambda}=\frac{(n-1)^2}{3n^2}\Big[3c^2+\gamma
\sigma^{-2} e^{-2 \lambda \phi}
\ln\{\sigma^{2} e^{2 \lambda \phi}\}+\beta
\sigma^{-2} e^{-2 \lambda \phi}\Big],
\end{eqnarray}
and
\begin{eqnarray}
w_{\Lambda }=-1-\frac{ 2\gamma
\sigma^{-2} e^{-2 \lambda \phi} -4\gamma
\sigma^{-2} e^{-2 \lambda \phi}\ln
(\sigma^{2} e^{2 \lambda \phi})-4\beta
\sigma^{-2} e^{-2 \lambda \phi}-6c^2}{3(3c^2+\gamma
\sigma^{-2} e^{-2 \lambda \phi}\ln
(\sigma^{2} e^{2 \lambda \phi})+\beta
\sigma^{-2} e^{-2 \lambda \phi})}\nonumber\\\times\left[1-\sqrt{\frac{3\Omega_\Lambda}{3c^2+\gamma
\sigma^{-2} e^{-2 \lambda \phi}\ln(\sigma^{2} e^{2 \lambda \phi})+\beta
\sigma^{-2} e^{-2 \lambda \phi}}}\right].\label{wh0}
\end{eqnarray}
The phantom crossing then occurs for the EoS parameter of the ECHDE
model in the following scenario \cite{rev}:
\begin{eqnarray}\label{c1}
2\gamma \sigma^{-2} e^{-2 \lambda \phi} -4\gamma
\sigma^{-2} e^{-2 \lambda \phi}\ln
(\sigma^{2} e^{2 \lambda \phi})-4\beta \sigma^{-2} e^{-2 \lambda \phi}
-6c^2>0,
\end{eqnarray}
\begin{eqnarray}\label{c2}
&3\Omega_\Lambda<3c^2+\gamma
\sigma^{-2} e^{-2 \lambda \phi}
\ln(\sigma^{2} e^{2 \lambda \phi})+\beta
\sigma^{-2} e^{-2 \lambda \phi}.
\end{eqnarray}
and
\begin{eqnarray}\label{c3}
2\gamma \sigma^{-2} e^{-2 \lambda \phi} -4\gamma
\sigma^{-2} e^{-2 \lambda \phi}\ln(
\sigma^{2} e^{2 \lambda \phi})-4\beta \sigma^{-2} e^{-2 \lambda \phi}
-6c^2<0,
\end{eqnarray}
\begin{eqnarray}\label{c4}
&3\Omega_\Lambda>3c^2+\gamma
\sigma^{-2} e^{-2 \lambda \phi}
\ln(\sigma^{2} e^{2 \lambda \phi})+\beta
\sigma^{-2} e^{-2 \lambda \phi}.
\end{eqnarray}
\begin{figure}[h]
\begin{center}
\includegraphics[scale=.3]{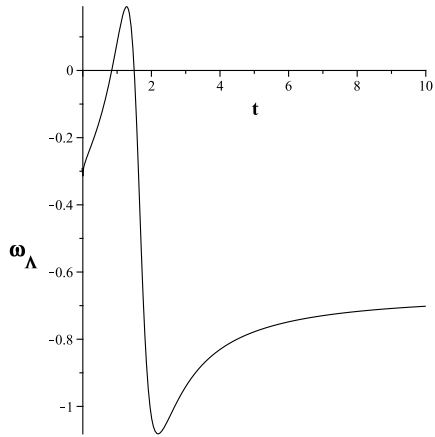}
\caption{Graph of the EoS parameter \((n > 0)\) in terms of time
evolution by choosing $c=0.5$, $\gamma=-2$, $n=2$ and
$\beta=0.25$.}\label{fig1}
\end{center}
\end{figure}

\begin{figure}[t]
\begin{center}
\includegraphics[scale=.3]{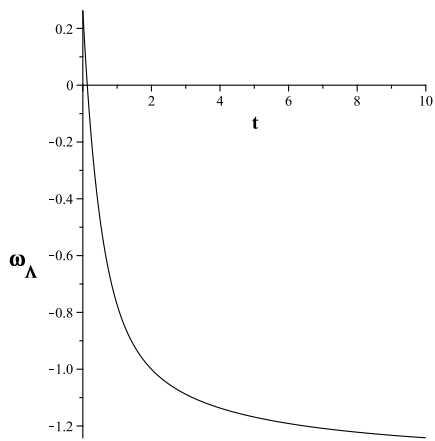}
\caption{Graph of the EoS parameter $(n < 0)$ in terms of time
evolution by choosing $c=3$, $\gamma=-2$, $n=-2$, $t_s=-1$ and
$\beta=2$.}\label{fig2}
\end{center}
\end{figure}
Using numerical calculation, the EoS parameter for both positive and
negative $n$ are given in the Figs. \ref{fig1} and \ref{fig2}. Also
we can see in both cases with different EoS parameters which the
phantom crossing occurs.

%################################################################################
\section{Conclusion}\label{5}
In this paper we started with scalar tensor theories with the non--minimal kinetic coupling to gravity and
obtained the corresponding field equations, energy density and pressure.
We introduced  the logarithmic entropy--corrected holographic
energy density as a dynamical cosmological constant. In order to obtain the cosmological parameters, we need to explicitly define the function $F(\phi)$. The popular exponential form is chosen with the motivation to produce phantom crossing behavior in formalism.\\ We obtained different
conditions in order to have
a correspondence between entropy--corrected holographic and
non-minimal kinetic coupling dark energy model. Also we reconstructed potential in terms of field $\phi$ for two cases $n > 0$ and $n< 0$. Finally we obtained the EoS parameter for the holographic
energy density in the model with the condition for phantom crossing scenario given by (\ref{c1})-(\ref{c4}).
%#####################################################################################
\section{Acknowledgements}\label{6}
The authors are indebted to the anonymous referee for his/her
comments that improved the Letter drastically.
%###############################################################################

\end{document}